\documentclass[a4paper]{jpconf}
\usepackage{graphicx}
\begin{document}
\title{Superconductivity without Local Inversion Symmetry; Multi-layer Systems}

\author{D Maruyama$^1$, M Sigrist$^2$ and Y Yanase$^1$}

\address{$^1$Department of Physics, Niigata University, Niigata
950-2181, Japan}
\address{$^2$Theoretische Physik, ETH-Honggerberg, 8093 Zurich, Switzerland}

\ead{marudai@phys.sc.niigata-u.ac.jp}

\begin{abstract}

While multi-layer systems can possess global inversion centers, they can have
regions with locally broken inversion symmetry.
This can modify the superconducting properties of such a system. 
Here we analyze two dimensional multi-layer systems yielding spatially modulated 
antisymmetric spin-orbit coupling (ASOC) and discuss superconductivity with mixed
parity order parameters. In particular, the influence of ASOC on the spin susceptibility
is investigated at zero temperature. For weak inter-layer coupling we find an enhanced
spin susceptibility induced by ASOC, which hints the potential importance of this aspect
for superconducting phase in specially structured superlattices.

\end{abstract}

\section{Introduction}

The discovery of superconductivity in the non-centrosymmetric
heavy Fermion compound CePt$_3$Si \cite{1,2}
has attracted much attention. Non-centrosymmetricity leads to antisymmetric
spin-orbit coupling (ASOC) which, for example, gives rise to mixed parity 
pairing, and the characteristic anisotropy of the spin susceptibility 
in the superconducting state [3-9].

Recently, artificial superlattices involving the heavy Fermion compound 
CeIn$_3$ \cite{10} and CeCoIn$_5$ \cite{11} have been fabricated, realizing
two-dimensional multi-layer structures of heavy electron systems and
ordinary metals. A particularly interesting case is the superlattice of 
CeCoIn$_5$ and YbCoIn$_5$ which shows superconductivity, most
likely induced by the CeCoIn$_5$ layers. 
Motivated by this system we setup a model of a superconductor with 
similar layered structure, including the fact that such structures show
distinctive violation of inversion (reflection) symmetry, while the overall
system has inversion centers. Such local non-centrosymmetricity is 
expected to have a pronounced effect on the superconducting phase
through local occurrence of ASOC, in particular, for the spin susceptibility.

\section{Formulation of model}

We introduce here model Hamiltonian for multi-layer systems including
ASOC, given by
\begin{eqnarray}
H&=&\sum_{{\bf k},s,m}\varepsilon({\bf k})c^{\dag}_{{\bf k}sm}c_{{\bf k}sm}+\sum_{{\bf k},s,s',m}\alpha_{m}{\bf g}({\bf k})\cdot{\bf \sigma}_{ss'}c^{\dag}_{{\bf k}sm}c_{{\bf k}s'm} \nonumber \\ &&+\frac{1}{2}\sum_{{\bf k},s,s',m}[\Delta_{ss'm}({\bf k})c^{\dag}_{{\bf k}sm}c^{\dag}_{-{\bf k}s'm}+\rm{h.c.}]+\sum_{{\bf k},{\it s},\langle {\it m,m'}\rangle}t_{\perp}c^{\dag}_{{\bf k}{\it sm}}c_{{\bf k}{\it sm'}},
\end{eqnarray}
where $c_{{\bf k}sm}$ ($c^{\dag}_{{\bf k}sm}$) is the annihilation
(creation) operator for an electron with spin $s$ on the layer $m$, 
and ${\bf \sigma}_{ss'}$ is the Pauli matrices. The $(x,y,z)$-axes correspond
to the $(a,b,c)$-axes of the tetragonal crystal.

The first two term describe the intra-layer dispersion and the
ASOC. We consider a square lattice with a tight-binding model, i.e.,
$\varepsilon({\bf k})=-2t(\cos{k_{x}}+\cos{k_{y}})-\mu$. 
We choose the unit of energy as $t=1$ and assume the chemical potential 
$\mu=-1$. The electron density per site is approximately 0.63.
The ASOC term preserves time reversal symmetry through the condition 
${\bf g}(-{\bf k})=-{\bf g}({\bf k})$ and has Rashba structure, which we assume a simple form
${\bf g}({\bf k})=(-\sin{k_{y}},\sin{k_{x}},0)$. The coupling constants
$ \alpha_m $ are layer dependent and have opposite sign for layers
above and below a center layer. 
The third term introduces intra-layer Cooper pairing via an off-diagonal mean field.
Here  $\Delta_{ss'm}({\bf k})$ involves both the spin singlet and triplet 
components, 
\begin{eqnarray}
\Delta_{ss'm}({\bf k})=\left(
\begin{array}{ccc}
-d_{xm}({\bf k})+{\rm{i}}d_{ym}({\bf k}) & \psi_{m}({\bf k})+d_{zm}({\bf k}) \\
-\psi_{m}({\bf k})+d_{zm}({\bf k}) & d_{xm}({\bf k})+{\rm{i}}d_{ym}({\bf k})
\label{op}
\end{array}
\right),
\end{eqnarray}
where $\psi_{m}({\bf k})$ and ${\bf d}_{m}({\bf k})$ are scalar and
vector order parameters for the spin-singlet and triplet pairing on
layer $m$, respectively. For simplicity we use an
order parameter on phenomenological grounds without resorting
to any microscopic model based on a pairing mechanism.
To be concrete we use an s-wave order parameter for the singlet 
and a p-wave order parameter for the triplet pairing, where on symmetry
grounds we request ${\bf d}_{m}({\bf k})\parallel{\bf g}({\bf k})$:
$\psi_{m}({\bf k})=\psi_{m}$ and 
${\bf d}_{m}({\bf k})=d_{m}{\bf g}({\bf
k})=d_{m}(-\sin{k_{y}},\sin{k_{x}},0)$. 
 We choose $|\psi_{m}|$, $|d_{m}| \leq 0.01$, small enough to
satisfy the condition
$|\Delta_{ss'm}({\bf k})|\ll|\alpha_{m}|\ll\varepsilon_{\rm F}$. 
The dominant order parameter component keeps the same sign 
over all layers, while the other (subdominant) component adjusts
the sign with the ASOC ($\alpha_{m}$).
The fourth term describes the inter-layer hopping of electrons between
nearest-neighbor layers. 
We assume that the inter-layer hopping $t_{\perp}$ is smaller 
than the intra-layer hopping $t$.

\section{Numerical results}

We now calculate the spin susceptibility of the multi-layer 
superconductors with spatially inhomogeneous ASOC, concentrating
on the magnetic field direction along the {\it c}-axis. 
The spin susceptibility 
$\chi={\rm lim}_{H \rightarrow 0} \langle M_s \rangle/ {H}$ 
is obtained numerically by calculating the magnetization $\langle M_s\rangle$
for a small magnetic field ${\bf H}$.  
The necessary Zeeman coupling term 
is given by,
\begin{eqnarray}
H_{\rm Z}=- \frac{g \mu_{\rm B}}{2} \sum_{{\bf k},s,s',m}{\bf H}\cdot{\bf
 \sigma}_{ss'}c^{\dag}_{{\bf k}sm}c_{{\bf k}s'm}, 
\end{eqnarray}
where $g = 2 $ and $\mu_{\rm B}$ is the Bohr magneton. 
First, the Hamiltonian is diagonalized in the presence of a field, introducing
the unitary transformation $ \hat{C}^{\dag}_{{\bf k}} = \hat{\Gamma}^{\dag}_{{\bf k}} \hat{U}^{\dag}({\bf k})$
 in Nambu-space of $ M $ layers, where the quasiparticle operators form a $4M$-dimensional vector
\begin{equation}
\hat{C}^{\dag}_{{\bf k}} = (c^{\dag}_{{\bf k} \uparrow 1} , c^{\dag}_{{\bf k} \downarrow 1}, c_{{- \bf k} \uparrow 1},  c_{{- \bf k} \downarrow 1}, \dots ,  c^{\dag}_{{\bf k} \uparrow M} , c^{\dag}_{{\bf k} \downarrow M}, c_{{- \bf k} \uparrow M}, c_{{- \bf k} \downarrow M})
\end{equation}
and analogous for the Bogoliubov quasiparticle operators $ \hat{\Gamma}^{\dag}_{{\bf k}} = (\gamma^{\dag}_{{\bf k} 1},  \gamma^{\dag}_{{\bf k} 2 }, \dots ,\gamma^{\dag} _{{\bf k}  4M})  $. Thus, the Hamiltonian is
\begin{equation}
H + H_{\rm Z} = \frac{1}{2}\sum_{{\bf k}}\sum_{i=1}^{4M} E_i({\bf k}) \gamma^{\dag}_{{\bf k} i} \gamma_{{\bf k} i} ,
\end{equation}
where $ E_i({\bf k}) $ are the quasiparticle energies. 
The magnetization is obtained as,
\begin{eqnarray}
\langle
 M_s \rangle=  \frac{g\mu_{\rm B}}{2}\sum_{{\bf k}}\sum^{4M}_{i=1}[\hat{S}^z ({\bf k})]_{ii}f(E_{i}({\bf k})),
\end{eqnarray}
where $ f(E) $ is the Fermi-Dirac distribution function. 
The matrix representation of spin operator is defined in the 
$\hat{\Gamma}^{\dag}_{{\bf k}}$ basis as
\begin{eqnarray}
\hat{S}^{\mu}({\bf k})=\hat{U}^{\dag}({\bf k}) \hat{\Sigma}^{\mu} \hat{U}({\bf k}),
\end{eqnarray}
with $\hat{\Sigma}^{\mu}$ the $\mu$-component of the spin operator 
in the $4M$-dimensional space. 

As concrete examples we discuss the spin susceptibility of bi-layer ($M=2$) 
and tri-layer ($M=3$) systems at $T=0$. 
The corresponding coupling constants of ASOC are described as 
$(\alpha_1, \alpha_2) = (\alpha, -\alpha)$ for bi-layers and 
$(\alpha_1, \alpha_2, \alpha_3) = (\alpha, 0, -\alpha)$ for tri-layers. 

\begin{figure}[h]
\begin{center}
\begin{minipage}{14pc}
\includegraphics[width=14pc]{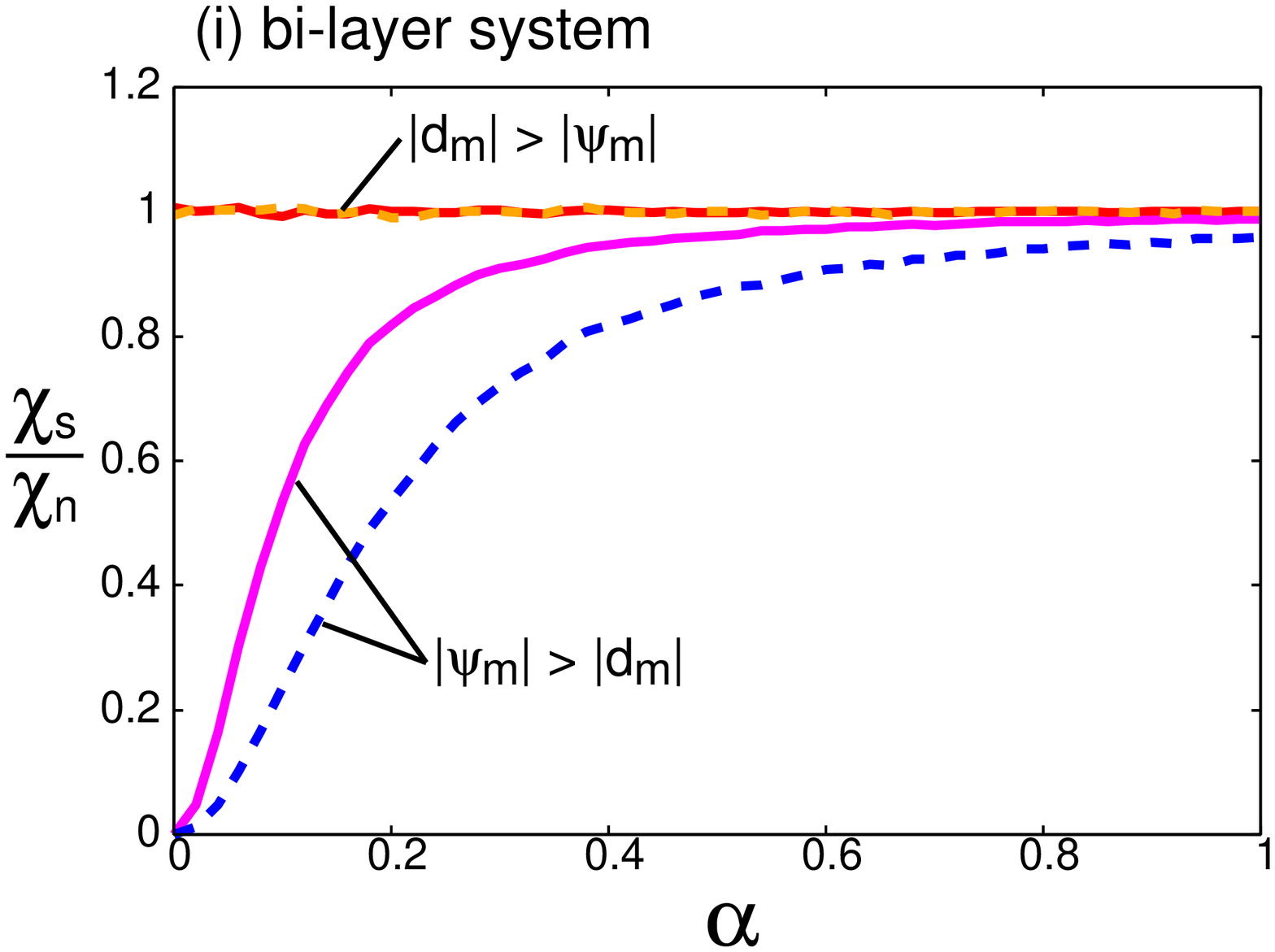}
\end{minipage}\hspace{2pc}%
\begin{minipage}{14pc}
\includegraphics[width=14pc]{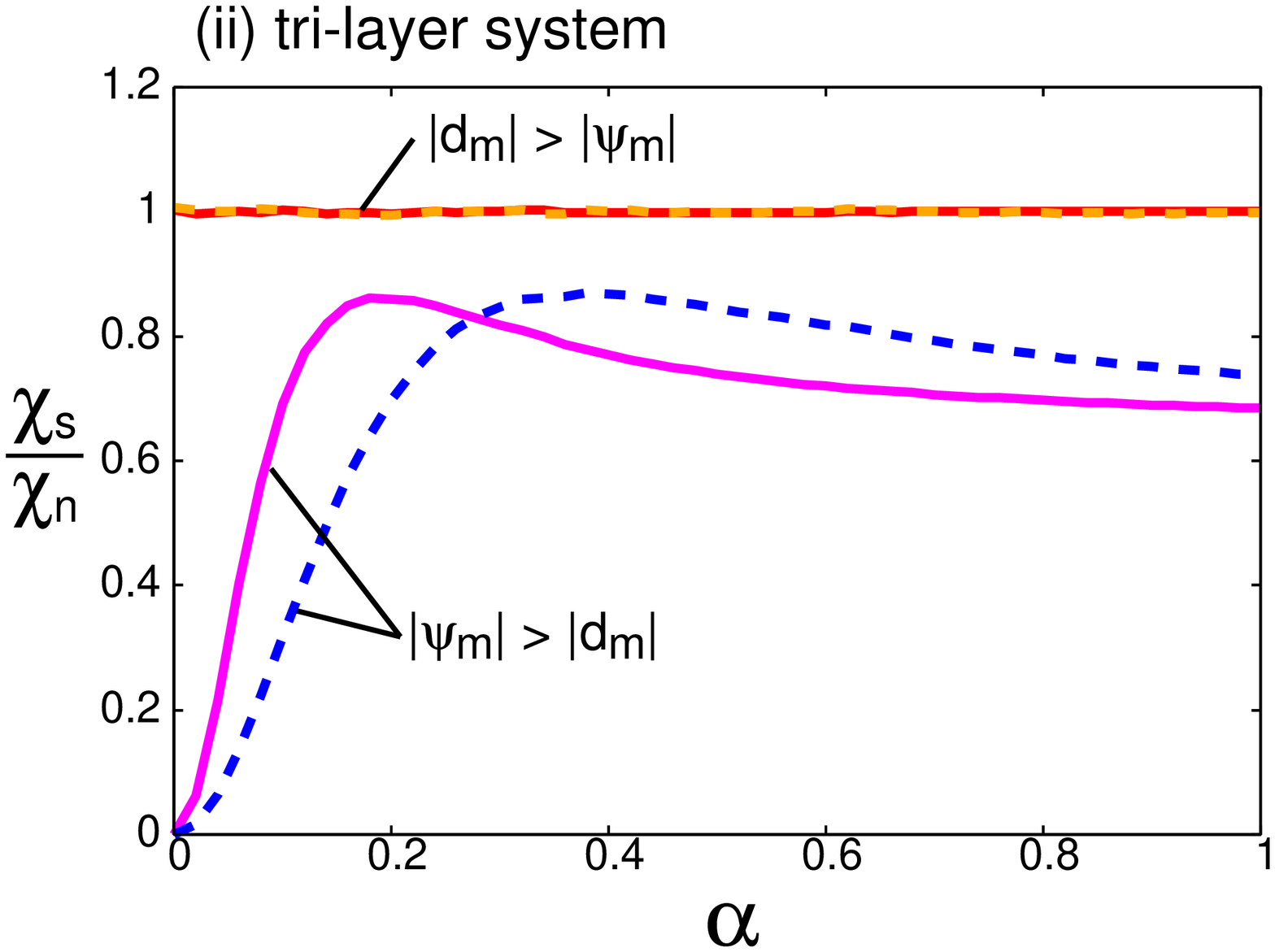}
\end{minipage} 
\caption{\label{m23}Spin susceptibility along {\it c}-axis for bi-layer system (i) and tri-layer system (ii). We assume $t_{\perp}=0.1$ (solid line) and $t_{\perp}=0.2$ (dashed line).}
\end{center}
\end{figure}

We compare now the two cases: (1) the spin triplet channel is dominant 
$|d_{m}|>|\psi_{m}|$ and (2) the spin singlet channel is dominant 
$|d_{m}|< |\psi_{m}|$. In case (1) the spin susceptibility remains 
unaffected by the superconducting state, $ \chi_{\rm s} = \chi_{\rm n} $, 
because the spin triplet component of the type $ {\bf d}_m ({\bf k})  
\propto {\bf g} ({\bf k}) \perp \hat{z} $ is an equal-spin pairing state 
with Cooper pair spins along the $c$-axis. Thus, spin polarization 
in the superconducting phase is possible without pair breaking. 
This feature is essentially independent of ASOC
and inter-layer hopping as can be seen in Fig.\ref{m23} for both 
the bi- and tri-layer systems. 

\begin{figure}[h]
\begin{center}
\includegraphics[width=14pc]{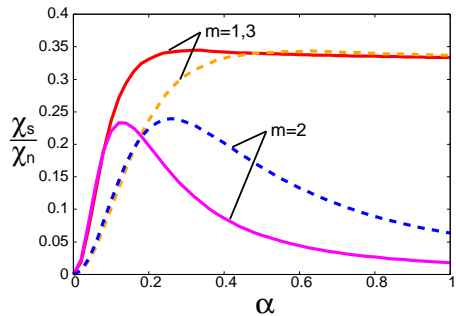}\hspace{2pc}%
\begin{minipage}[b]{14pc}\caption{\label{el}Spin susceptibility along 
{\it c}-axis for tri-layer system. The contribution from each layer is shown. 
We assume $|\psi_{m}|>|d_{m}|$, and $t_{\perp}=0.1$ (solid line) and 
$t_{\perp}=0.2$ (dashed line).}
\end{minipage}
\end{center}
\end{figure}

More interesting is case (2) as spin singlet pairing leads to complete
suppression of the spin susceptibility at $T=0$ in a conventional superconductor.
Indeed for vanishing ASOC ($\alpha=0$) we find
$ \chi_{\rm s} = 0 $ irrespective of $ t_{\perp} $. As soon as 
ASOC is turned on, however, the spin susceptibility
gradually recovers and approaches for large $ \alpha $ a constant
value:  $ \chi_{\rm s} \to \chi_{\rm n} $ for the bi-layer and 
$ \chi_{\rm s} \to 2 \chi_{\rm n}/3 $ for the tri-layer. The mechanism for
this behavior lies in the spin-splitting of the
electronic spectrum due to the Rashba-type ASOC. 
The spin polarization involves now an inter-band Van-Vleck-type
contribution which is only weakly affected by the opening of
quasiparticle gap in the superconducting phase. Note that this
inter-band contribution relies on the ASOC for
matrix elements with the Zeeman coupling and is only available
for the layers with non-vanishing $ \alpha $. Consequently,
in the bi-layer system all layers are involved, giving rise to
full recovery of $ \chi_{\rm s} $ for large $ \alpha $ (analogous to the
uniformly non-centrosymmetric superconductor \cite{5}), while in the
tri-layer system only two of three layers can contribute
yielding a correspondingly reduced limiting value for $ \chi_{\rm s} $. 
Figure~\ref{el} corroborates this picture by considering the contributions
of the different layers. Indeed in the large $\alpha $ regime the outer 
layers $ m= 1,3 $ carrying 
ASOC saturate at $ \chi_{\rm s} \to  \chi_{\rm n} /3 $
while the center layer $ m =2 $ completely suppresses. Remarkably
at small  $ \alpha $ ($ < t_{\perp}$), $\chi_{\rm s} $ behaves for all layers
in the same way and leads for the center layer to a striking non-monotonic
$ \alpha $-dependence. 

The numerical data in Fig.\ref{m23} show that the inter-layer hopping is 
in competition with ASOC, such that a larger $ t_{\perp} $
yields a higher effective $ \alpha_{\rm eff} \sim t_{\perp} $ for the crossover
from the behavior of conventional superconductor to that of non-centrosymmetric superconductor.
This crossover is best evident in the peak of $ \chi_{\rm s} $ around $  \alpha_{\rm eff} \sim t_{\perp} $
for the center layer (Fig.\ref{el}). 
Thus, modifying $ t_{\perp} $, e.g., by uniaxial stress along the {\it c}-axis,
can influence the magnetic response for {\it c}-axis fields in case (2).
No such effect is expected for case (1).  

Within our model we find that the spin susceptibility along 
the {\it ab}-axis is always the half of the value observed along {\it c}-axis, 
independent of the strength of $ \alpha $ and $ t_{\perp} $ and  
the number of layers. Furthermore, we find that the spin susceptibility
for both field directions is affected by the phase difference of 
order parameter between layers, but independent of the ratio of 
spin singlet and triplet components. 
Details will be explained elsewhere.

\section{Conclusion}

In view of the fact that CeCoIn$_5$ is known to realize spin singlet superconductivity, 
we believe that most likely case (2) of our discussion is relevant for the multi-layer systems. 
Thus, the large observed upper critical fields in the superlattice of CeCoIn$_5$ \cite{11}
would then rely on the presence of the spatially modulated ASOC. Moreover we believe that the 
variability of the superlattices and
also the possibility of local measurements of magnetic properties through NMR
would give many intriguing insights into the aspect of ASOC in these 
artificial systems. 

\ack

The authors are grateful to H. Shishido, T. Shibauchi, Y. Matsuda, and M. Fischer
for fruitful discussions. 
This work was supported by a Grant-in-Aid for Scientific Research 
on Innovative Areas ``Heavy Electrons'' (No. 21102506) from MEXT, Japan. 
It was also supported by a Grant-in-Aid for 
Young Scientists (B) (No. 20740187) from JSPS. We are also grateful for
financial support of the Swiss Nationalfonds and the NCCR MaNEP.

\end{document}